\documentclass{article}

\usepackage[preprint]{neurips_2025}

\usepackage[utf8]{inputenc}
\usepackage[T1]{fontenc}
\usepackage{url}
\usepackage{booktabs}
\usepackage{amsfonts}
\usepackage{amsmath}
\usepackage{amssymb}
\usepackage{nicefrac}
\usepackage{microtype}
\usepackage{graphicx}
\usepackage{xcolor}
\usepackage{colortbl}
\usepackage{hyperref}
\usepackage{cleveref}
\crefname{appendix}{appendix}{appendices}
\Crefname{appendix}{Appendix}{Appendices}
\usepackage{multirow}
\usepackage{array}
\usepackage{listings}
\usepackage{xspace}

\hypersetup{colorlinks=true,citecolor=blue,linkcolor=blue,urlcolor=blue}

\lstdefinestyle{promptstyle}{
  basicstyle=\ttfamily\footnotesize,
  breaklines=true,
  columns=fullflexible,
  keepspaces=true,
  showstringspaces=false,
  literate={—}{{--}}2 {–}{{--}}1 {→}{{$\to$}}1
}

\newcommand{\method}{\textsc{RISE}\xspace}
\newcommand{\methodb}{\textsc{RISE-BM25}\xspace}

\title{Towards Retrieving Interaction Spaces for \\ Agentic Search}

\author{%
Shengyao Zhuang\thanks{Corresponding author: \texttt{shengyaozhuang@gmail.com}}\\
\And
Yuansheng Ni\\
\And
Hengxin Fun
\And
Jimmy Lin\\
\And
Xueguang Ma
}

\date{}

\begin{document}

\maketitle

\begin{abstract}
Retrieval for search agents is still inherited from non-agentic information
retrieval: a retriever ranks the corpus and the agent reads a small set of
returned documents. Recent direct corpus interaction (DCI) work shows that agents can
instead \emph{interact} with the raw corpus through shell tools such as
\texttt{grep} and file reads. But
unbounded interaction does not scale: every broad shell command is a scan
over the whole corpus, and latency degrades sharply as the corpus grows.
We argue that the role of retrieval for agentic search is not just to
select documents that fit in the LLM context window, but to construct an
\emph{interaction space}: a bounded subset of the corpus the agent can
explore with associated tools. Two design consequences follow. The space needs a \emph{boundary} supplied by retrieval, and the
objects within it should be \emph{processed for interaction}. As a
proof of concept, we propose \textsc{RISE} (\emph{Retrieving
Interaction SpacE}): we use BM25 to construct the interaction space;
meanwhile, its documents are processed during indexing for shell-style
navigation. On BrowseComp-Plus, \textsc{RISE} matches the pure-shell
DCI baseline at $78\%$ accuracy with \texttt{gpt-5.4-mini} at roughly
one quarter of the per-query cost. At 1M documents,
\methodb{} reaches $81\%$ on
\texttt{gpt-5.4-mini}, whereas DCI on
\texttt{gpt-5.4-nano} degrades to $60\%$ with $33$ of $100$
wall-clock failures.
\end{abstract}

\begin{center}
\href{https://github.com/texttron/RISE}{%
\raisebox{-0.15em}{\includegraphics[height=0.95em]{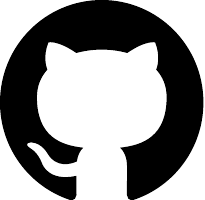}}\;%
\texttt{github.com/texttron/RISE}}
\end{center}

\begin{figure}[!htbp]
\centering
\includegraphics[width=\textwidth]{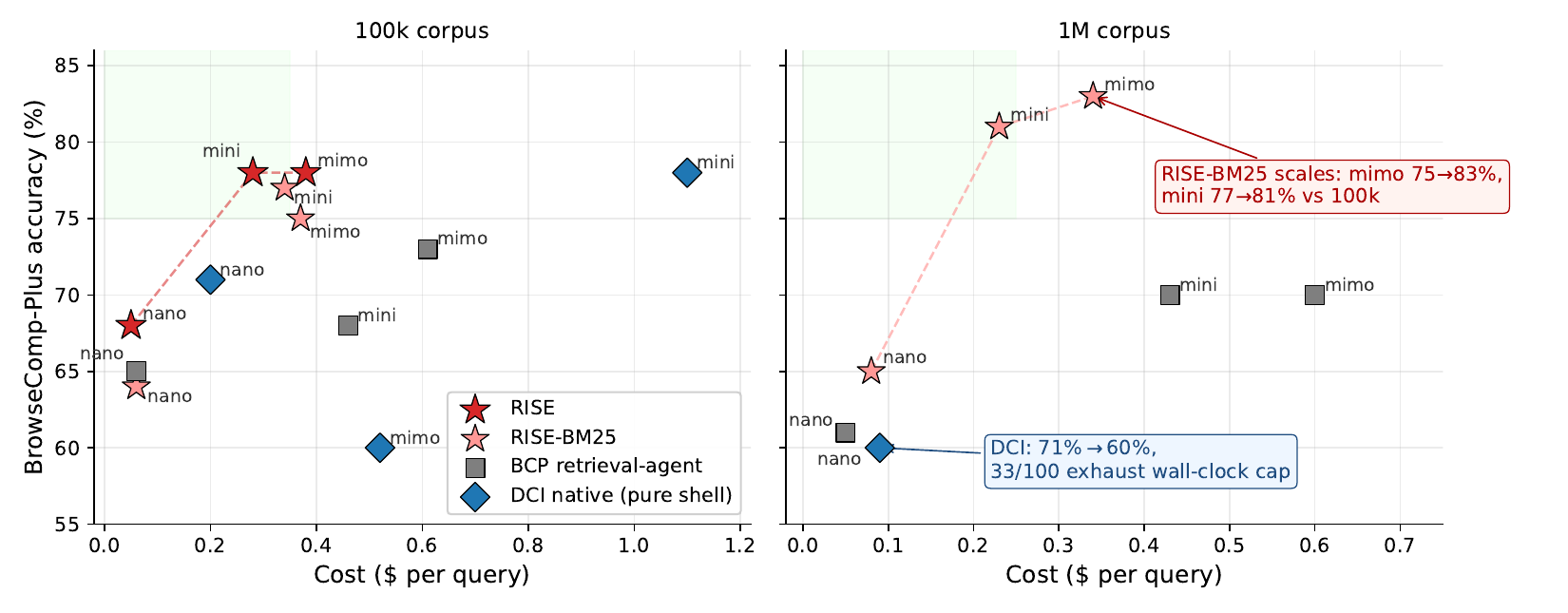}

\caption{\textbf{Accuracy vs.\ per-query cost on BrowseComp-Plus.}
Each point is one model--architecture pair; the dashed line is the
\method{} frontier. At 100k, \method{} matches DCI at 78\%
accuracy on \emph{mini} while reducing cost from \$1.10 to \$0.28/query. At 1M, \methodb{} (the bounding mechanism alone) remains stable after adding 900k FineWeb-Edu distractors; the
1M DCI run is reported for \emph{nano}, where accuracy drops to 60\% with
33/100 wall-clock failures.}
\label{fig:teaser}
\end{figure}

\section{Introduction}
\label{sec:intro}

Retrieval for search agents has, until recently, inherited its interface
from non-agentic information retrieval. At each turn, a retriever ranks the
corpus and returns a small set of documents (or snippets cut from them),
and the agent answers from this context, just as a non-agentic reader
would. This contract underlies retrieval-augmented
generation~\citep{lewis2020rag}, browser-assisted question
answering~\citep{nakano2021webgpt}, and ReAct-style search
agents~\citep{yao2023react}. Retrieval's job is snippet selection: pick
which short fragments will fit in the agent's prompt this turn.

A growing line of work breaks this contract from the agent side. Direct
corpus interaction (DCI)~\citep{li2026dci} gives the agent shell tools
over the raw corpus filesystem; coding-agent work on long-context
processing represents large text collections as files manipulated with
native tools~\citep{cao2026codingagents}; and harness studies show that
lexical search inside an agent loop is a strong primitive when the model
can iteratively issue and refine shell commands~\citep{sen2026grep}. The
shift is from reading pre-extracted context to interacting with the
corpus; the agent no longer consumes a fixed evidence window, it
explores.

But unbounded interaction does not scale. DCI's own corpus-scaling study
reports that on a 100-query BrowseComp-Plus subset, expanding the corpus
from 100k to 200k documents increases average tool calls from $38.5$ to
$86.9$, more than doubles latency and cost, and drops accuracy by $13.6$
points; at 400k documents, accuracy falls to $37.5\%$ and $20$ of $100$
queries hit the maximum tool budget~\citep{li2026dci}. The cost is
structural: every broad \texttt{grep} is a scan over the whole corpus, so
as the candidate space widens, the agent spends its budget triaging shell
output rather than answering. Snippet retrieval and unbounded shell
access fail in opposite directions (one loses evidence resolution, the
other loses scale), but they share an assumption about what retrieval is
\emph{for}: that its job is to decide what the agent reads next.

We argue for a different framing. For agentic search, retrieval should
not just select documents that fit in the LLM context window, but
construct an \emph{interaction space}: a bounded subset of the corpus
the agent can explore with associated tools. Two design consequences
follow. Each is stated together with what it rules out, so neither
reduces to a tautology. \emph{(i) The space needs a boundary supplied
by retrieval.} The boundary is neither the absence of a candidate set
(DCI's full-corpus shell), nor the context window itself (snippet
retrieval, where evidence must fit in the prompt), nor an implicit
ranking (a list the agent cannot re-traverse with shell tools), but an
explicit persistent set the agent operates on outside the context
window. \emph{(ii) The objects in the space should be processed for
interaction, not stored as raw text.} The form of processing is
neither raw text (which forces a full scan for any fact), nor
retrieval-time chunks (which pre-decide what evidence the agent sees),
nor summaries (which lose the verbatim grounding the agent needs for
evidence-checking), but in-place structural metadata that lets shell
tools jump to relevant spans without document-scale reads. To this end,
we propose \method{} (\emph{Retrieving Interaction SpacE}): we use
BM25 to construct the interaction space, and process its documents
during indexing for shell-style navigation. \methodb{} is the
boundary-only ablation that implements only (i).

On BrowseComp-Plus, \method{} matches the pure-shell DCI baseline at
$78\%$ accuracy with \texttt{gpt-5.4-mini} at one quarter of the
per-query cost. Removing the document-processing step (i.e., \methodb{}
at 100k) drops accuracy by $1$--$4$ points across model tiers, isolating
the contribution of consequence (ii) at the scale where it is evaluated.
When the corpus is expanded $10\times$ to 1M documents, \methodb{}
remains stable, while DCI degrades to $60\%$ with $33$ of $100$
wall-clock failures.

\section{Related Work}
\label{sec:related}

\paragraph{Retrieval-augmented and deep research agents.}
RAG introduced the now-standard pattern of coupling a parametric model
with an external retrieval index~\citep{lewis2020rag}. WebGPT and ReAct
made the interaction loop explicit: the model issues actions, observes
external information, and continues reasoning~\citep{nakano2021webgpt,yao2023react}.
Recent deep research systems extend this loop to longer horizons, where
agents repeatedly search, inspect evidence, and synthesize answers or
reports. Search-R1 trains language models to generate multi-turn search
queries during reasoning~\citep{jin2025searchr1}, while WebThinker
interleaves reasoning, web navigation, and report drafting for large
reasoning models~\citep{li2025webthinker}. BrowseComp and
BrowseComp-Plus provide hard browsing-style evaluations for this regime,
with BrowseComp-Plus replacing opaque live web search by a fixed corpus
that makes retrieval choices comparable~\citep{wei2025browsecomp,chen2025browsecompplus}.
Our work uses the same agentic search setting but studies a different
question: what corpus interface should the agent receive after retrieval
has selected an initial evidence set?

\paragraph{Retrieval and ranking for agentic search.}
A growing line of work adapts retrieval models to the behavior of search
agents rather than treating agent queries as ordinary ad-hoc queries.
AgentIR argues that deep research agents expose useful reasoning traces
before search calls, and trains a retriever to embed the trace together
with the query~\citep{chen2026agentir}. Agentic-R trains retrievers for
multi-turn agentic search using both local passage relevance and global
answer correctness~\citep{liu2026agenticr}. Complementarily,
\citet{meng2026revisiting} analyze text ranking in deep research on
BrowseComp-Plus, comparing retrieval units, retrievers, re-rankers, and
query transformations. SAGE studies retrieval for scientific deep
research agents and finds that keyword-oriented agent queries can make
strong lexical baselines competitive with dense
retrieval~\citep{hu2026sage}.
These works improve or diagnose the ranking component, treating
retrieval's output as a ranked list whose top entries are delivered into
the agent's context. \method{} is orthogonal: it keeps BM25 deliberately
simple and changes what retrieval \emph{returns}, namely a bounded set of
candidate documents that lives outside the context window, is traversable
with the agent's shell tools, and whose objects are themselves processed
for shell-style navigation.

\paragraph{Corpus interfaces for agentic search.}
Recent work on corpus interfaces for agents falls into three classes
that differ in what retrieval returns to the agent. The first class
delivers a ranked list of snippets into the context window, as in
conventional RAG and most retrieval-agent pipelines, where
\citet{sen2026grep} show that the success of retrieval inside an agent
loop depends strongly on the harness and on how retrieved information
is delivered to the model. The second class removes the candidate
boundary entirely and exposes the raw corpus to shell tools:
DCI~\citep{li2026dci} argues directly for terminal-level interaction
with the full collection, and \citet{cao2026codingagents} reach a
similar interface from the long-context side, treating large text
collections as file systems that coding agents manipulate with native
tools. The third class returns a bounded set of candidate documents
that the agent operates on with shell tools outside the context
window; PI-SERINI~\citep{hsu2026piserini} is the closest existing
instance: like \method{}, it uses BM25 to bound the candidate set, but
exposes that set as a cached ranking the agent pages through rather
than as a navigable workspace. \method{} sits in this third class as a
filesystem-workspace variant: retrieval returns a bounded,
re-explorable \emph{interaction space} whose objects carry in-place
structural metadata so shell tools can jump to sub-document evidence,
rather than a cached ranking within the same class or the interfaces
of the first two classes.

\section{Method}
\label{sec:method}

We operationalize the two consequences from \Cref{sec:intro} as the
two layers of \method{}:

\begin{itemize}
\item[(i)] \textbf{Bounded workspace.} The interaction space needs an
explicit, persistent set of candidate documents that lives outside the
model's context window and is re-traversable with shell tools.
Retrieval supplies this set; the agent operates inside it.
\item[(ii)] \textbf{Navigable objects.} The documents placed in that
set should be processed at indexing time so that shell tools can reach
sub-document evidence without scanning the whole file.
\end{itemize}

\begin{figure}[t]
\centering
\includegraphics[width=\textwidth]{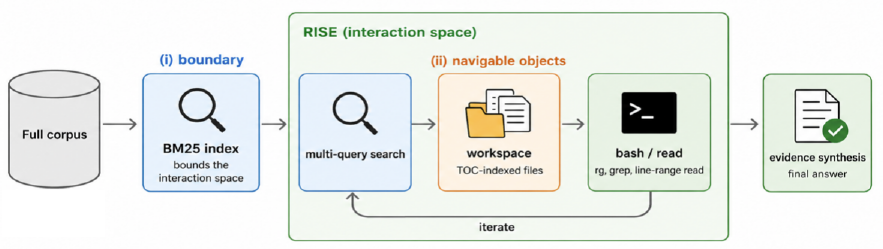}
\caption{\textbf{\method{} workflow.} BM25 supplies the boundary
(consequence i): candidate files are imported from the full corpus into a
per-query interaction space that lives outside the context window. The
imported files are TOC-indexed by an offline processing step (consequence
ii): each file exposes a line-numbered table of contents so shell tools
can land on relevant spans without a full read. The agent then iterates
between multi-query search, shell inspection, and targeted reads inside
the bounded space before synthesizing a final answer.}
\label{fig:rise_overview}
\end{figure}

\noindent \method{} instantiates both; \methodb{} is the boundary-only
ablation that implements only (i). \Cref{fig:rise_overview} shows the
overall workflow, and \Cref{sec:method_bounding} and
\Cref{sec:method_processing} develop the two consequences in turn, each
pinned down by what it rules out.

\subsection{Bounding the Interaction Space}
\label{sec:method_bounding}

Consequence (i) is a claim about \emph{what retrieval returns}: an
explicit, persistent set of candidate documents that lives outside the
model's context window and is re-traversable with shell tools. This
rules out three alternatives. It is not the absence of a candidate set
(as in DCI's full-corpus shell, where every broad command scans the
whole corpus). It is not a context-window boundary (as in snippet
retrieval, where the agent only sees what already fits in the prompt).
And it is not an implicit ranking (a list the agent cannot re-traverse
with native tools after a single read). \methodb{} instantiates exactly
this: a per-query working directory the agent treats as a filesystem.

We populate the directory $\mathcal{W}$ with multi-query BM25 retrieval.
The agent issues \texttt{search(queries)} with one or more
natural-language sub-queries; each sub-query retrieves the top $K$
documents from a corpus-wide BM25 index, and the union is hardlinked
into $\mathcal{W}$, preserving corpus-relative paths. \texttt{search}
is the only tool that can import documents, and $\mathcal{W}$ grows
monotonically across the conversation. The observation returned to the
model is a top-10 preview per sub-query, but the full top-$K$ matches
are available to subsequent shell and read calls, so the agent can use
search as a high-recall import operation rather than a narrow snippet
reader. Inside $\mathcal{W}$, the agent uses \texttt{bash} (running
standard utilities such as \texttt{rg}, \texttt{grep}, \texttt{cat})
and \texttt{read} (returning line-numbered slices of a file) to
inspect and verify imported documents. Implementation budgets (output
truncation, subprocess timeouts, tool-call cap) are reported in
\Cref{sec:setup}; exact prompts are in
\Cref{app:plain_prompt,app:structured_docs}.

The mechanism is one of many that would satisfy (i): any retriever
returning a re-explorable candidate set would do, including dense,
late-interaction, or hybrid alternatives. We keep BM25 deliberately
simple to isolate the interface from retriever engineering.

\subsection{Processing Objects for Interaction}
\label{sec:method_processing}

Consequence (ii) is a claim about how objects inside the interaction
space should be \emph{processed for interaction}, not how they should
be ranked or chunked. This too rules out three alternatives. It is not
raw text (which forces the agent to scan a document end-to-end for any
fact it might contain). It is not retrieval-time chunking or snippet
delivery (where the indexing pipeline has already decided what slice of
the document the model is allowed to see, collapsing the interaction
back into snippet retrieval, forfeiting the workspace, and preventing
the agent from re-exploring the document for evidence the chunker
discarded). And it is not summarization (which loses the verbatim text
the agent needs to commit an answer). What it is, instead, is in-place
structural metadata: the object is rewritten so that shell tools can
land directly on relevant sub-document spans, while the original body
text remains available for verification.

\method{} uses one form of in-place metadata: a line-numbered table of
contents (TOC) with section anchors prepended to the body text.
Documents are processed offline once with a two-stage pipeline. A
section-proposal LLM (\texttt{gpt-5.4-nano} at low reasoning effort, via the OpenAI Batch API)
emits JSON section titles, descriptions, and verbatim anchor strings;
a deterministic postprocessor validates each anchor by exact substring
match (or short whitespace-normalized fallback) and inserts validated
anchors as \texttt{\#\#} headings together with a
\texttt{\#\# Table of Contents} block listing each section's
1-indexed line range. Body text is never summarized, deleted, or
rewritten; the original document is preserved for verification. The
pipeline is one-time and corpus-scale (about \$0.0014 per document at
our 100k scale); statistics, an example output, and full prompts are
in \Cref{app:structured_docs,app:structured_example,app:structured_generation_prompt}.
At inference time, \texttt{search} hardlinks the structured
counterpart of each retrieved document, and the agent prompt
instructs it to use the TOC for navigation but verify any committed
fact against the body text.

Other forms of in-place metadata would also satisfy (ii): hierarchical
section graphs, paragraph-level anchors, or learned navigation hints
share the requirement of letting shell tools reach sub-document
evidence without document-scale reads. We study TOCs because they are
the cheapest in-place form that supports both broad browsing (via the
TOC) and precise line-range reads (via the section anchors).

\section{Experimental Setup}
\label{sec:setup}

\paragraph{Dataset and corpora.}
All experiments use BrowseComp-Plus~\citep{chen2025browsecompplus}, a
fixed-corpus version of difficult browsing questions. Each run requires a full interactive agent trajectory followed by LLM-as-judge evaluation, which is costly in both API spend and wall-clock time; under this budget we evaluate on a fixed sample of 100 queries from the released BrowseComp-Plus question set.
The 100k setting uses the released
BrowseComp-Plus corpus. The 1M setting keeps the same questions and gold
documents, but adds 900,000 fixed-seed FineWeb-Edu distractors
\citep{penedo2024fineweb} to the 100k corpus.

\paragraph{Agent models and judging.}
We evaluate three agent models: Xiaomi's \texttt{mimo-v2.5-pro}, OpenAI's
\texttt{gpt-5.4-mini} with medium reasoning effort, and
\texttt{gpt-5.4-nano} with high reasoning effort. All answers are judged
with an LLM-as-judge protocol that follows the official BrowseComp-Plus
Appendix~F prompt, using \texttt{gpt-5.1} as the judge model. We report
judged accuracy, no-answer counts, mean tool calls, and mean
agent cost per query in USD. We also report two evidence-coverage
diagnostics: \emph{gold\_R}, whether gold documents appear in the BM25
candidate union, and \emph{cov\_mean}, whether gold documents are
surfaced to the agent through reads or shell output. The latter is useful
for diagnosing each architecture, but is not directly comparable across
all systems because their tool interfaces expose evidence differently.

\paragraph{Execution environment.}
All interactive agent runs were executed on a MacBook Air with an Apple
M4 chip and 24GB of unified memory. We ran each method separately, rather
than co-scheduling different systems, and used four workers within a run,
so at most four queries were active concurrently.

\paragraph{Compared systems.}
We compare \method{} (the full system from \Cref{sec:method}, with
both BM25 bounding and offline TOC processing), \methodb{} (the
boundary-only ablation, identical to \method{} except that the
workspace contains the original documents instead of their
TOC-augmented counterparts), and two baselines. Both \method{} and
\methodb{} use bm25s~\citep{lu2024bm25s} defaults ($k_1{=}1.5,b{=}0.75$),
retrieve $K{=}1000$ documents per sub-query, and surface a top-10
preview while letting the agent inspect the accumulated workspace with
\texttt{bash} and line-range \texttt{read}; we use $K{=}1000$ as the
default because the resulting working sets are usually on the order of,
or below, ten thousand unique files, at which scale local shell
operations remain fast and changing $K$ has little direct effect on
model cost. \method{} is evaluated only on the 100k corpus, since
extending the offline TOC pipeline to the added 900k FineWeb-Edu
distractors would require a one-time preprocessing pass we did not
incur; this is a preprocessing-budget choice, not a limitation of the
interface. The \emph{retrieval-agent} baseline reimplements the
BrowseComp-Plus Appendix~E agent with BM25 search at $k{=}5$, 512-token
snippets, and a \texttt{get\_document} tool. \emph{DCI} is a
direct-corpus filesystem baseline with \texttt{bash} and \texttt{read}
over the full corpus, no retriever, the DCI prompt, and the
DCI-Agent-Lite \texttt{level3} runtime context-management setting,
which applies tool-result truncation and compaction.

\paragraph{Budgets and stopping conditions.}
Each run has both a model-call budget and a wall-clock cap. For the
reproduced baselines, we follow the budget and stopping-condition settings
from their original papers unless stated otherwise. \method{} and \methodb{} (and the retrieval-agent baseline) use
100 model calls and a one-hour per-query wall-clock cap. DCI uses 300 model calls and a 1.5-hour per-query
wall-clock cap, following its original configuration; full-corpus shell
scans are I/O-bound and need more turns. This asymmetry deliberately
favours DCI---it receives $3\times$ the model calls and $1.5\times$ the
wall-clock budget of \method{}---so any cost or robustness gap we report
in DCI's disfavour is a conservative one.
We disable forced final-answer coercion at the cap for all headline runs:
if an agent exhausts its model-call or wall-clock budget without
committing an answer, the run is recorded as a no-answer case with the
corresponding stopping condition. For shell tools, both \method{} and \methodb{} truncate bash output
to 4000 characters and cap each subprocess at 60 seconds;
DCI uses the DCI-style bash truncation of the last 2000 lines or 50KB.
For document inspection, both \method{} and \methodb{} use line-range
reads with a 2000-line default; DCI reads are capped by the DCI-style head truncation at 2000
lines or 50KB, and the retrieval-agent \texttt{get\_document} tool
returns full document text as in the BrowseComp-Plus baseline.

\section{Results}
\label{sec:experiments}

\begin{table*}[t]
\centering
\small
\setlength{\tabcolsep}{4pt}
\caption{Cross-model results on the 100-query BrowseComp-Plus sample
with the 100k corpus. Cost is USD per query; within each model tier,
\textbf{bold} marks the best accuracy and the lowest cost. \emph{Turns} is the mean number
of model (API) calls per query. \emph{no\_ans}
counts queries with no final answer; superscripts indicate the stopping
condition: $^c$ context length, $^t$ turn/API-call cap, $^w$ wall-clock
cap, and $^o$ other non-answer stop. \emph{Search}, \emph{Bash}, and \emph{Read} are mean tool calls per
query; for the retrieval-agent, \emph{Read} is \texttt{get\_document}.
\emph{gold\_R} is mean gold-document recall in the retrieved candidate set
and \emph{cov\_mean} is mean gold-document coverage of the agent's
inspected context; both are evidence-coverage diagnostics and are not
directly comparable across architectures. The final \texttt{gpt-5.4} row
is a state-of-the-art upper-bound probe, not part of the main
cross-architecture comparison.}
\label{tab:headline}
\resizebox{\textwidth}{!}{%
\begin{tabular}{llcccccccccc}
\toprule
Model & Architecture & Acc $\uparrow$ & no\_ans & Turns & Search & Bash & Read & $\text{gold}_R$ & cov\_mean & \$/q $\downarrow$ \\
\midrule
\multirow{4}{*}{mimo-v2.5-pro}
  & \method{}        & $\textbf{78\%}$ & $2^{t}$           & $23.5$ & $16.2$ & $4.8$  & $7.9$ & $79.4\%$ & $80.2\%$ & $0.38$ \\
  & \methodb{}               & $75\%$          & $1^{t}$           & $19.8$ & $15.7$ & $3.6$  & $6.5$ & $75.7\%$ & $75.7\%$ & $\textbf{0.37}$ \\
  & retrieval-agent         & $73\%$          & $1^{t}$           & $34.5$ & $30.8$ & ---    & $2.5$ & $76.5\%$ & $76.5\%$ & $0.61$ \\
  & DCI                     & $60\%$          & $18^{w}$          & $26.0$ & ---    & $51.8$ & $5.5$ & ---      & $62.0\%$ & $0.52$ \\
\midrule
\multirow{4}{*}{\shortstack[l]{gpt-5.4-mini\\(medium effort)}}
  & \method{}        & $\textbf{78\%}$ & $1^{c}$           & $24.3$ & $13.1$ & $9.2$  & $6.4$ & $78.5\%$ & $79.6\%$ & $\textbf{0.28}$ \\
  & \methodb{}               & $77\%$          & $1^{w}$           & $23.0$ & $14.8$ & $9.6$  & $5.2$ & $77.7\%$ & $77.9\%$ & $0.34$ \\
  & retrieval-agent         & $68\%$          & $12^{c,o}$        & $29.2$ & $37.0$ & ---    & $1.9$ & $77.5\%$ & $77.5\%$ & $0.46$ \\
  & DCI                     & $\textbf{78\%}$ & $5^{w,t}$         & $48.8$ & ---    & $90.3$ & $8.8$ & ---      & $82.4\%$ & $1.10$ \\
\midrule
\multirow{4}{*}{\shortstack[l]{gpt-5.4-nano\\(high effort)}}
  & \method{}        & $68\%$          & $0$               & $24.2$ & $10.4$ & $8.4$   & $4.6$ & $73.4\%$ & $74.8\%$ & $\textbf{0.05}$ \\
  & \methodb{}               & $64\%$          & $0$               & $22.9$ & $10.2$ & $8.1$   & $3.7$ & $72.2\%$ & $75.4\%$ & $0.06$ \\
  & retrieval-agent         & $65\%$          & $0$               & $17.0$ & $14.3$ & ---     & $2.0$ & $62.0\%$ & $62.0\%$ & $0.06$ \\
  & DCI                     & $\textbf{71\%}$ & $7^{w}$           & $45.7$ & ---    & $119.4$ & $7.1$ & ---      & $77.4\%$ & $0.20$ \\
\midrule

\shortstack[l]{gpt-5.4}
  & \method{}        & $82\%$ & $8^{t}$           & $32.2$ & $11.0$ & $16.2$  & $7.1$ & $88.5\%$ & $92.4\%$ & $1.25$ \\
\bottomrule
\end{tabular}}
\end{table*}
\subsection{Headline Results}

\Cref{tab:headline} compares \method{}, the \methodb{} ablation, and the
two baselines on the 100k corpus. \method{} is the best or tied-best system on
the two larger model tiers: it reaches $78\%$ accuracy with
\texttt{mimo-v2.5-pro} and \texttt{gpt-5.4-mini}, while costing less than
the retrieval-agent and DCI baselines in both cases. \methodb{} is close
behind, indicating that most of the gain comes from bounding the interaction space; the indexing-time document processing provides a smaller additional improvement.

The pure-shell DCI baseline is competitive when it has enough model and
wall-clock budget: it ties \method{} on \texttt{gpt-5.4-mini} and is the
best nano-tier system. Its cost profile is different, however. We give DCI
a deliberately generous budget---$3\times$ the model calls and $1.5\times$
the wall-clock cap of \method{} (\Cref{sec:setup})---yet on
\texttt{gpt-5.4-mini} it still costs $4\times$ more per query to reach the
same accuracy, and on \texttt{mimo-v2.5-pro} it exhausts even that larger
budget on many queries ($18$ of $100$ wall-clock failures) and falls to
$60\%$ accuracy. The asymmetry favours the baseline, so the efficiency gap
is a conservative estimate: bounding the interaction space is enough that
\method{} needs only a third of DCI's call budget, while unbounded shell
access cannot reliably finish within three times as much. The retrieval-agent baseline is cheaper
than DCI but consistently below \method{} on the larger models, despite
similar BM25 gold-document recall. This suggests that its main limitation
is not finding candidate documents, but exposing enough of them to the
agent.

To probe the ceiling of the interface with a state-of-the-art model, we
also run \method{} with the full \texttt{gpt-5.4} medium reasoning effort (last row of
\Cref{tab:headline}). It is the most accurate configuration overall at
$82\%$, and surfaces gold documents into context most reliably
($\text{cov\_mean}{=}92.4\%$, roughly ten points above the strongest
smaller-tier configuration at $82.4\%$), confirming that the interaction-space
interface keeps paying off as model capability grows. The gain over \method{} on \texttt{gpt-5.4-mini} is
modest ($+4$ points) and comes at $4.4\times$ the per-query cost
($\$1.25$ vs.\ $\$0.28$): on harder queries the stronger model exhausts
its turn budget rather than committing a low-confidence guess ($8$
turn-cap no-answers, the most of any cell). The accuracy--cost frontier
therefore still favours the small-model tiers; \texttt{gpt-5.4} marks the
upper bound the same interface reaches when cost is no object.

\subsection{Corpus Scaling}
\label{sec:scaling}

\Cref{tab:scaling} evaluates the corpus-portable systems at 1M documents. At this scale we evaluate only \methodb{} (the bounding mechanism without document processing), since extending the offline document-processing pipeline to the added 900k FineWeb-Edu distractors would require a one-time preprocessing pass we did not incur. \textbf{The 1M numbers thus speak to the first design consequence (boundary) only; the second consequence's behavior at scale is not directly measured here.} This is a preprocessing-cost constraint, not an interface constraint. We also run DCI at 1M only for the nano tier,
where DCI is the strongest nano-agent system at 100k. The larger-model
DCI settings already show substantial wall-clock pressure at 100k, making
1M runs less informative as comparisons of agent behavior; even the nano
1M DCI run took roughly two days on our machine.

\methodb{} does not degrade under the 10$\times$ corpus expansion,
and modestly improves: accuracy goes from $75\%$ to $83\%$ on
\texttt{mimo-v2.5-pro}, from $77\%$ to $81\%$ on \texttt{gpt-5.4-mini}, and
from $64\%$ to $65\%$ on \texttt{gpt-5.4-nano}. 
We do not over-interpret these gains. Adding 900k distractors shifts the corpus idf statistics, which can make the default BM25 parameters better suited to the agent's search queries; it is also possible that some of the added documents are relevant but not annotated as gold in BrowseComp-Plus. Either way, the key point is robustness: accuracy does not degrade under a 10$\times$ corpus expansion, and \methodb{}'s shell exploration stays confined to the retrieved working set rather than the full 1M files.

The snippet retrieval-agent is less robust. It drops on the mimo and nano
tiers and remains below \methodb{} in every 1M comparison. DCI shows the
opposite scaling failure: on nano, accuracy drops from $71\%$ to $60\%$,
and wall-clock failures rise from 7 to 33 queries. In the 1M setting, the
agent often spends its budget triaging broad shell outputs before it can
commit to an answer. The lower reported DCI API cost at 1M is therefore
not an efficiency gain: many queries spend more wall-clock time waiting
for full-corpus \texttt{bash} commands, reach the wall-clock cap after
fewer model interactions, and consequently incur fewer billed model
tokens despite worse accuracy.

\begin{table*}[t]
\centering
\caption{Scaling from 100k to 1M documents on the 100-query BrowseComp-Plus sample.
All \method{}-family rows at 1M are \methodb{} (the boundary-only configuration), since the offline TOC pipeline was not extended to the 900k FineWeb-Edu distractors. Tool-call columns report 1M values
with the change from 100k in parentheses. \emph{Read*} is the
\texttt{read} count for \methodb{} and the \texttt{get\_document} count for
the retrieval-agent. DCI is reported at 1M only
for \texttt{gpt-5.4-nano}.}
\label{tab:scaling}
\resizebox{\textwidth}{!}{%
\begin{tabular}{llcccccccc}
\toprule
 & & \multicolumn{2}{c}{Accuracy} & \multicolumn{4}{c}{Tool calls / query} & \\
\cmidrule(lr){3-4} \cmidrule(lr){5-8}
Model & Architecture & 100k & 1M ($\Delta$) & Turns & Search & Bash & Read* & \$/q 1M \\
\midrule
\multirow{2}{*}{mimo-v2.5-pro}
  & \textbf{\methodb{}}     & $75\%$ & $\textbf{83\%}\,(+8)$ & $18.7\,(-1.1)$ & $14.5\,(-1.2)$ & $2.6\,(-1.0)$  & $6.8\,(+0.3)$ & $0.34$ \\
  & retrieval-agent         & $73\%$ & $70\%\,(-3)$          & $35.1\,(+0.6)$ & $31.7\,(+0.9)$ & ---            & $2.2\,(-0.3)$ & $0.60$ \\
\midrule
\multirow{2}{*}{\shortstack[l]{gpt-5.4-mini\\(medium effort)}}
  & \textbf{\methodb{}}     & $77\%$ & $\textbf{81\%}\,(+4)$ & $22.6\,(-0.4)$ & $15.6\,(+0.8)$ & $6.9\,(-2.7)$  & $5.5\,(+0.3)$ & $0.23$ \\
  & retrieval-agent         & $68\%$ & $70\%\,(+2)$          & $28.5\,(-0.7)$ & $32.4\,(-4.6)$ & ---            & $1.7\,(-0.2)$ & $0.43$ \\
\midrule
\multirow{3}{*}{\shortstack[l]{gpt-5.4-nano\\(high effort)}}
  & \textbf{\methodb{}}     & $64\%$ & $\textbf{65\%}\,(+1)$  & $25.1\,(+2.2)$  & $10.6\,(+0.4)$ & $9.7\,(+1.6)$   & $3.9\,(+0.2)$ & $0.08$ \\
  & retrieval-agent         & $65\%$ & $61\%\,(-4)$           & $16.7\,(-0.3)$  & $14.0\,(-0.3)$ & ---             & $2.2\,(+0.2)$ & $0.05$ \\
  & DCI                     & $71\%$ & $60\%\,(-11)$          & $24.2\,(-21.5)$ & ---            & $70.3\,(-49.1)$ & $4.5\,(-2.6)$ & $0.09$ \\
\bottomrule
\end{tabular}}
\end{table*}

\begin{table}[t]
\centering
\footnotesize
\setlength{\tabcolsep}{5pt}
\caption{BM25 top-$K$ ablation for \methodb{} on the 100k
corpus. $K$ is applied per sub-query. \emph{WS size} is the mean number of
unique files in the accumulated workspace at the end of a query.}
\label{tab:ablation_k}
\begin{tabular}{lccccccc}
\toprule
Model & $K$ & Acc $\uparrow$ & \$/q $\downarrow$ & Turns & Bash & cov\_mean & WS size \\
\midrule
\multirow{3}{*}{mimo-v2.5-pro}
  & $100$                   & $\textbf{76\%}$ & $\textbf{0.350}$ & $19.7$ & $3.4$ & $76.1\%$ & $1{,}069$ \\
  & $1000$ \emph{(default)} & $75\%$          & $0.372$          & $19.8$ & $3.6$ & $75.7\%$ & $7{,}590$ \\
  & $10000$                 & $73\%$          & $0.367$          & $20.4$ & $4.1$ & $77.8\%$ & $41{,}821$ \\
\midrule
\multirow{3}{*}{\shortstack[l]{gpt-5.4-mini\\(medium effort)}}
  & $100$                   & $75\%$          & $0.337$          & $25.9$ & $10.2$ & $78.2\%$ & $1{,}334$ \\
  & $1000$ \emph{(default)} & $\textbf{77\%}$ & $0.337$          & $23.0$ & $9.6$  & $77.9\%$ & $9{,}121$ \\
  & $10000$                 & $75\%$          & $\textbf{0.308}$ & $25.4$ & $9.7$  & $82.0\%$ & $45{,}603$ \\
\midrule
\multirow{3}{*}{\shortstack[l]{gpt-5.4-nano\\(high effort)}}
  & $100$                   & $64\%$          & $0.064$          & $23.2$ & $9.5$ & $71.9\%$ & $1{,}441$ \\
  & $1000$ \emph{(default)} & $64\%$          & $0.060$          & $22.9$ & $8.1$ & $75.4\%$ & $10{,}363$ \\
  & $10000$                 & $\textbf{65\%}$ & $\textbf{0.057}$ & $22.6$ & $8.2$ & $76.5\%$ & $49{,}640$ \\
\bottomrule
\end{tabular}
\end{table}

\subsection{BM25 Top-$K$ Ablation}
\label{sec:ablation_k}

\Cref{tab:ablation_k} varies the number of BM25 documents hardlinked per
sub-query. Increasing $K$ expands the final working directory, but it
does not monotonically improve accuracy. The default $K{=}1000$ is the
best setting for \texttt{gpt-5.4-mini}, within one point of the best
setting for \texttt{mimo-v2.5-pro}, and tied on nano up to one point. In
practice this setting yields accumulated workspaces of roughly
$7.6$k--$10.4$k files, a range where local shell operations remain fast on
our evaluation machine. The smaller $K{=}100$ setting is competitive and
more compact, while $K{=}10000$ produces $42$k--$50$k-file workspaces
without clear accuracy gains.

This ablation supports the main design choice: \method{} does not need
to expose the agent to as much of the corpus as possible. It needs a
working set large enough to contain relevant documents from multiple
sub-queries, but small enough for shell exploration to remain local.
Because additional BM25 hits are added as local files rather than injected
into the model context, changing $K$ has little direct effect on API cost;
the main tradeoff is recall versus local navigation overhead.

\begin{figure*}[t]
\centering
\includegraphics[width=\textwidth]{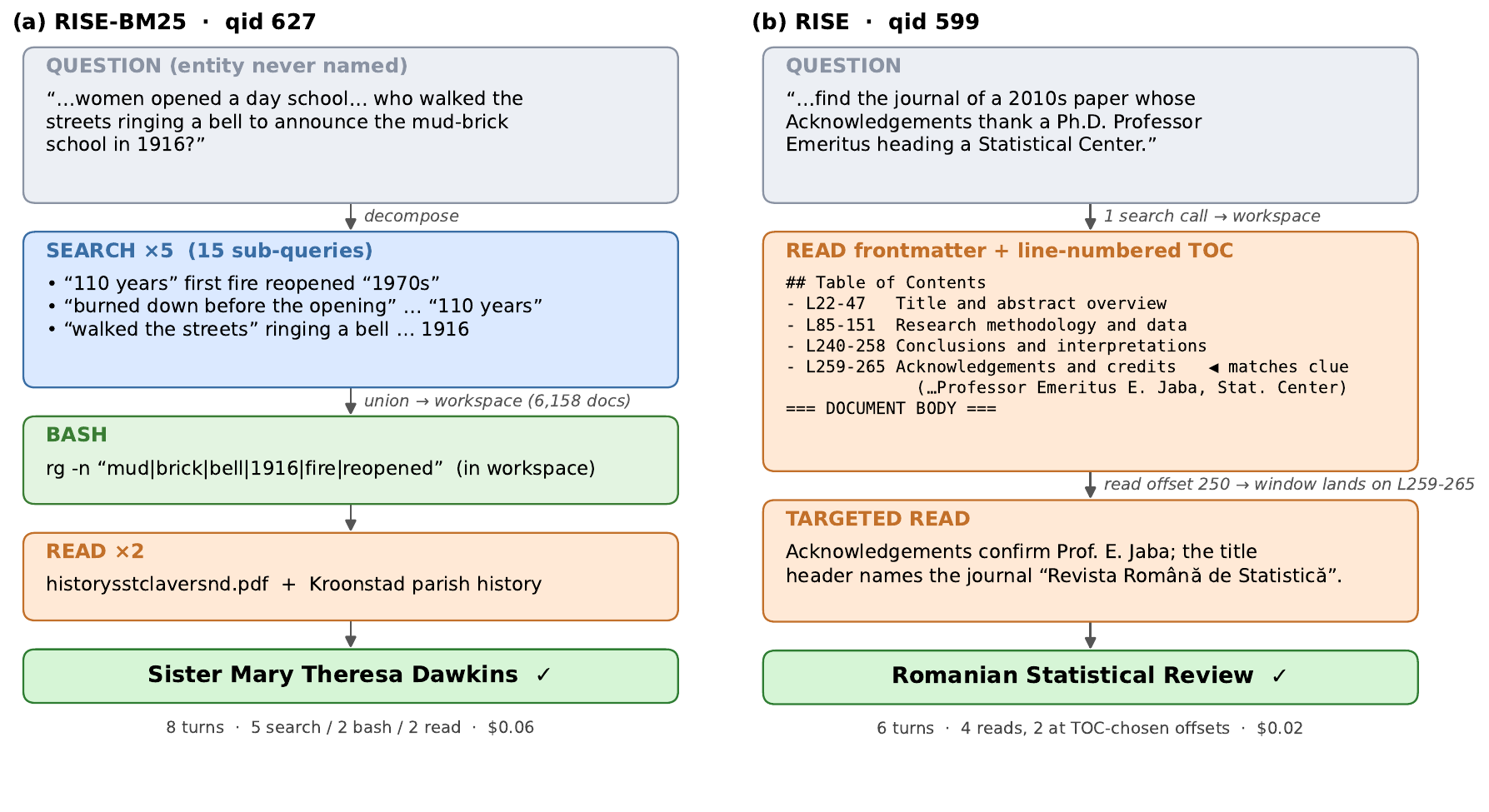}
\caption{Two case-study trajectories (\texttt{gpt-5.4-mini}, 100k
corpus). \textbf{(a) \methodb{}}: an obfuscated question is
decomposed into many BM25 sub-queries whose union forms the
workspace, after which \texttt{bash}/\texttt{read} verify the answer
in the workspace. \textbf{(b) \method{}}: the agent reads a
document's frontmatter and line-numbered table of contents, matches the
``Acknowledgements'' clue to the TOC entry \texttt{L259--265:
Acknowledgements and credits}, and jumps to that line range to confirm
the answer (reading only the relevant spans rather than the whole
document).}
\label{fig:case}
\end{figure*}

\section{Case Studies}
\label{sec:case}

\Cref{fig:case} traces two correct \texttt{gpt-5.4-mini} trajectories on
the 100k corpus in which the agent follows the intended workflow end to
end. Both queries are obfuscated, multi-clue BrowseComp-Plus questions
whose answer entity is never named.

In the \methodb{} case (qid 627), the agent never knows the target entity, so
it cannot retrieve it directly. Instead it decomposes the question into
fifteen paraphrased sub-queries across five \texttt{search} calls (each
locking onto a different clue cluster), and accumulates their BM25 unions
into one workspace; only then does it switch to in-workspace
verification with \texttt{rg} and \texttt{read}. This is the intended
division of labour: retrieval repeatedly \emph{shapes the workspace},
and the shell \emph{verifies within it}.

In the \method{} case (qid 599), the agent opens a candidate paper and
is shown its LLM-generated, line-numbered table of contents. Rather than
read the full body, it maps a clue to a TOC entry (the question asks
about the paper's \emph{Acknowledgements}, and the entry
\texttt{L259--265: Acknowledgements and credits} names exactly the
Professor Emeritus and Statistical Center the clue describes), then
issues a targeted \texttt{read} at that line range to confirm the paper,
then reads the title header for the journal name. Two of its four
reads are issued at non-zero offsets chosen from the table of contents:
exactly the navigation pattern that indexing-time document processing
(consequence ii) is designed to enable.

Read together, the two trajectories visualize the two consequences
side by side: retrieval shaping a workspace for in-place verification,
and indexing-time processing turning long documents into navigable
structure.

\section{Conclusion}
\label{sec:conclusion}

We argued that for agentic search, retrieval should construct an
\emph{interaction space} rather than extract context, with two design
consequences: the space needs a boundary supplied by retrieval, and its
objects should be processed for interaction. \method{} is a
proof-of-concept instantiation: BM25 bounds the space, and documents in
it are offline-augmented with line-numbered tables of contents to make
them shell-navigable.

On BrowseComp-Plus, this interface produces a favorable accuracy--cost
tradeoff: \method{} matches the pure-shell DCI baseline at a fraction
of its per-query cost. \methodb{} remains stable when the corpus
expands from 100k to 1M documents, whereas pure-shell DCI degrades
sharply with wall-clock failures, and the snippet retrieval-agent
remains weaker on the larger model tiers.

The broader conclusion is that agentic search systems need an
intermediate retrieval output between a ranked snippet list and the
unrestricted corpus filesystem. An interaction space provides that
output: a bounded set of candidates the agent can re-explore with shell
tools (the boundary consequence), populated with objects whose
structure makes sub-document evidence cheap to reach (the processing
consequence). This makes corpus interaction a design problem in
\emph{what retrieval returns} and \emph{how the returned objects are
structured}, not only in retriever quality or agent prompting. It also
raises the question of what makes a returned interaction space
\emph{good}: a question that present retrieval benchmarks, built for
the older contract, were not designed to answer.

\section{Limitations}
\label{sec:limitations}

Our experiments instantiate each design consequence in a single
concrete way: BM25 for the boundary mechanism, and line-numbered TOCs
for the indexing-time document processing. Other retrievers (dense,
late-interaction, hybrid) and other in-place metadata schemes
(section graphs, paragraph-level anchors) are natural follow-ups
that would help characterize how broadly the framing holds. The
processing consequence is evaluated only on the 100k corpus, because
rerunning the offline TOC pipeline on the 900k FineWeb-Edu
distractors used at 1M was outside our budget; the 1M scaling result
therefore speaks primarily to the boundary consequence. A fully
crossed design that includes the fourth corner (snippet retrieval
over the TOC-augmented corpus) would more cleanly separate the
contribution of the filesystem-workspace interface from that of BM25
prefiltering, and is a useful next step.

Beyond these design-space gaps, the evaluation itself is narrow in scope: a single benchmark family
(BrowseComp-Plus), 100 queries, closed-weight agents, and a single
judge (\texttt{gpt-5.1}) sharing a model family with two of three
agents and the structured-corpus generator. As a result, per-tier
accuracy gaps of a few points are best read as directional rather
than precise.

\section*{Acknowledgements}
We thank \href{https://justram.github.io/}{Jheng-Hong Yang} for helpful
discussions.

\bibliographystyle{plainnat}
\bibliography{references}

@inproceedings{lewis2020rag,
author = {Lewis, Patrick and Perez, Ethan and Piktus, Aleksandra and Petroni, Fabio and Karpukhin, Vladimir and Goyal, Naman and K\"{u}ttler, Heinrich and Lewis, Mike and Yih, Wen-tau and Rockt\"{a}schel, Tim and Riedel, Sebastian and Kiela, Douwe},
title = {Retrieval-augmented generation for knowledge-intensive NLP tasks},
year = {2020},
isbn = {9781713829546},
publisher = {Curran Associates Inc.},
address = {Red Hook, NY, USA},
booktitle = {Proceedings of the 34th International Conference on Neural Information Processing Systems},
articleno = {793},
numpages = {16},
location = {Vancouver, BC, Canada},
series = {NIPS '20}
}

@article{nakano2021webgpt,
  title   = {{WebGPT}: Browser-assisted Question-answering with Human Feedback},
  author  = {Nakano, Reiichiro and Hilton, Jacob and Balaji, Suchir and Wu, Jeff and Ouyang, Long and Kim, Christina and Hesse, Christopher and Jain, Shantanu and Kosaraju, Vineet and Saunders, William and Jiang, Xu and Cobbe, Karl and Eloundou, Tyna and Krueger, Gretchen and Button, Kevin and Knight, Matthew and Chess, Benjamin and Schulman, John},
  journal = {arXiv preprint arXiv:2112.09332},
  year    = {2021}
}

@inproceedings{yao2023react,
  title     = {{ReAct}: Synergizing Reasoning and Acting in Language Models},
  author    = {Yao, Shunyu and Zhao, Jeffrey and Yu, Dian and Du, Nan and Shafran, Izhak and Narasimhan, Karthik and Cao, Yuan},
  booktitle = {International Conference on Learning Representations},
  year      = {2023}
}

@inproceedings{
jin2025searchr1,
title={{Search-R1}: Training {LLM}s to Reason and Leverage Search Engines with Reinforcement Learning},
author={Bowen Jin and Hansi Zeng and Zhenrui Yue and Jinsung Yoon and Sercan O Arik and Dong Wang and Hamed Zamani and Jiawei Han},
booktitle={Second Conference on Language Modeling},
year={2025},
url={https://openreview.net/forum?id=Rwhi91ideu}
}

@inproceedings{
li2025webthinker,
title={{WebThinker}: Empowering Large Reasoning Models with Deep Research Capability},
author={Xiaoxi Li and Jiajie Jin and Guanting Dong and Hongjin Qian and Yongkang Wu and Ji-Rong Wen and Yutao Zhu and Zhicheng Dou},
booktitle={The Thirty-ninth Annual Conference on Neural Information Processing Systems},
year={2026},
url={https://openreview.net/forum?id=7LKKHBAMzH}
}

@article{chen2026agentir,
  title   = {{AgentIR}: Reasoning-Aware Retrieval for Deep Research Agents},
  author  = {Chen, Zijian and Ma, Xueguang and Zhuang, Shengyao and Lin, Jimmy and Asai, Akari and Zhong, Victor},
  journal = {arXiv preprint arXiv:2603.04384},
  year    = {2026}
}

@article{liu2026agenticr,
  title   = {{Agentic-R}: Learning to Retrieve for Agentic Search},
  author  = {Liu, Wenhan and Ma, Xinyu and Zhu, Yutao and Li, Yuchen and Shi, Daiting and Yin, Dawei and Dou, Zhicheng},
  journal = {arXiv preprint arXiv:2601.11888},
  year    = {2026}
}

@article{meng2026revisiting,
  title   = {Revisiting Text Ranking in Deep Research},
  author  = {Meng, Chuan and Ou, Litu and MacAvaney, Sean and Dalton, Jeff},
  journal = {arXiv preprint arXiv:2602.21456},
  year    = {2026}
}

@article{hu2026sage,
  title   = {{SAGE}: Benchmarking and Improving Retrieval for Deep Research Agents},
  author  = {Hu, Tiansheng and Zhao, Yilun and Zhang, Canyu and Cohan, Arman and Zhao, Chen},
  journal = {arXiv preprint arXiv:2602.05975},
  year    = {2026}
}

@article{hsu2026piserini,
  title   = {Rethinking Agentic Search with {PI-SERINI}: Is Lexical Retrieval Sufficient?},
  author  = {Hsu, Tz-Huan and Yang, Jheng-Hong and Lin, Jimmy},
  journal = {arXiv preprint arXiv:2605.10848},
  year    = {2026}
}

@article{wei2025browsecomp,
  title   = {{BrowseComp}: A Simple Yet Challenging Benchmark for Browsing Agents},
  author  = {Wei, Jason and Sun, Zhiqing and Papay, Spencer and McKinney, Scott and Han, Jeffrey and Fulford, Isa and Chung, Hyung Won and Passos, Alex Tachard and Fedus, William and Glaese, Amelia},
  journal = {arXiv preprint arXiv:2504.12516},
  year    = {2025}
}

@article{chen2025browsecompplus,
  title   = {{BrowseComp-Plus}: A More Fair and Transparent Evaluation Benchmark of Deep-Research Agent},
  author  = {Chen, Zijian and Ma, Xueguang and Zhuang, Shengyao and Nie, Ping and Zou, Kai and Liu, Andrew and Green, Joshua and Patel, Kshama and Meng, Ruoxi and Su, Mingyi and Sharifymoghaddam, Sahel and Li, Yanxi and Hong, Haoran and Shi, Xinyu and Liu, Xuye and Thakur, Nandan and Zhang, Crystina and Gao, Luyu and Chen, Wenhu and Lin, Jimmy},
  journal = {arXiv preprint arXiv:2508.06600},
  year    = {2025}
}

@article{li2026dci,
  title   = {Beyond Semantic Similarity: Rethinking Retrieval for Agentic Search via Direct Corpus Interaction},
  author  = {Li, Zhuofeng and Zhang, Haoxiang and Wei, Cong and Lu, Pan and Nie, Ping and Lu, Yi and Bai, Yuyang and Feng, Shangbin and Zhu, Hangxiao and Zhong, Ming and Zhang, Yuyu and Xie, Jianwen and Choi, Yejin and Zou, James and Han, Jiawei and Chen, Wenhu and Lin, Jimmy and Jiang, Dongfu and Zhang, Yu},
  journal = {arXiv preprint arXiv:2605.05242},
  year    = {2026}
}

@article{cao2026codingagents,
  title   = {Coding Agents are Effective Long-Context Processors},
  author  = {Cao, Weili and Yin, Xunjian and Dhingra, Bhuwan and Zhou, Shuyan},
  journal = {arXiv preprint arXiv:2603.20432},
  year    = {2026}
}

@article{sen2026grep,
  title   = {Is Grep All You Need? How Agent Harnesses Reshape Agentic Search},
  author  = {Sen, Sahil and Kasturi, Akhil and Lumer, Elias and Gulati, Anmol and Subbiah, Vamse Kumar},
  journal = {arXiv preprint arXiv:2605.15184},
  year    = {2026}
}

@article{lu2024bm25s,
  title   = {{BM25S}: Orders of Magnitude Faster Lexical Search via Eager Sparse Scoring},
  author  = {L{\`u}, Xing Han},
  journal = {arXiv preprint arXiv:2407.03618},
  year    = {2024}
}

@inproceedings{penedo2024fineweb,
  title     = {The {FineWeb} Datasets: Decanting the Web for the Finest Text Data at Scale},
  author    = {Penedo, Guilherme and Kydl{\'i}{\v{c}}ek, Hynek and Ben Allal, Loubna and Lozhkov, Anton and Mitchell, Margaret and Raffel, Colin and von Werra, Leandro and Wolf, Thomas},
  booktitle = {Advances in Neural Information Processing Systems, Datasets and Benchmarks Track},
  year      = {2024}
}

\clearpage
\appendix
\crefalias{section}{appendix}
\crefalias{subsection}{appendix}
\section{\methodb{} System Prompt}
\label{app:plain_prompt}

The \methodb{} runs use the following system prompt. The user prompt
is the BrowseComp-Plus query under a one-line \texttt{QUESTION:} header,
shared across \method{} and \methodb{}.

\begin{lstlisting}[style=promptstyle]
You answer research questions over a large document corpus you can't see directly.

You have three tools:
- search(queries): search the corpus with one or more queries in a single call. Each query is matched as a bag of words against document text, so write queries as natural-language descriptions with several distinctive terms; pass multiple complementary queries together for broader coverage. Returns a per-query top-10 preview with file paths and short snippets — the preview is only a sample; the full match set per query (hundreds to thousands of docs) is ADDED to your working directory and accumulates across turns, so use bash/read to explore beyond the preview.
- bash(command): run a shell command (rg, grep, ls, find, cat, head, etc.) over your working directory. Paths are relative to the working directory; absolute paths (anything starting with `/`) won't find anything — bash can't see outside the working directory. Use rg/grep to find phrases across retrieved docs. Files are plain text; run `ls` first to see the directory layout.
- read(file_path): read a file by its full path.

Use the tools iteratively: search to pull evidence into your working directory, bash/read to inspect it, and repeat until confident. When confident, output the final answer in this format and stop calling tools:
Explanation: {your reasoning}
Exact Answer: {your succinct final answer}
Confidence: {0-100%}
\end{lstlisting}

\section{Structured Documents: Prompt and Example}
\label{app:structured_docs}

\method{} uses the same search and shell interface as \methodb{}, but
the system prompt additionally explains the TOC-augmented file format
and instructs the agent to verify facts in the body rather than in TOC
summaries.

\paragraph{Generation statistics (100k corpus).} The offline pipeline
processed all $100{,}195$ documents with no API or postprocessing
failures. It produced at least one validated TOC entry for $94.5\%$ of
documents, and located $802{,}177$ of $808{,}062$ proposed section
anchors by exact or whitespace-normalized substring match ($99.3\%$).
Of the rest, $22$ documents had all proposed anchors fail validation and
$4$ produced unparseable model output; these fall back to the original
text with an empty TOC. Section proposal used \texttt{gpt-5.4-nano} at
low reasoning effort.

\subsection{\method{} System Prompt (with Document Processing)}
\begin{lstlisting}[style=promptstyle]
You answer research questions over a large document corpus you can't see directly.

You have three tools:
- search(queries): search the corpus with one or more queries in a single call. Each query is matched as a bag of words against document text, so write queries as natural-language descriptions with several distinctive terms; pass multiple complementary queries together for broader coverage. Returns a per-query top-10 preview with file paths and short snippets — the preview is only a sample; the full match set per query (hundreds to thousands of docs) is ADDED to your working directory and accumulates across turns, so use bash/read to explore beyond the preview.
- bash(command): run a shell command (rg, grep, ls, find, cat, head, etc.) over your working directory. Paths are relative to the working directory; absolute paths (anything starting with `/`) won't find anything — bash can't see outside the working directory. Use rg/grep to find phrases across retrieved docs. Files are plain text; run `ls` first to see the directory layout.
- read(file_path, offset?, limit?): read a file by its full path. Optional 0-indexed `offset` and `limit` (both in LINES, not chars) let you read a slice instead of the whole file; defaults to lines 0..2000.

Every document begins with a YAML frontmatter block (title / author / date), followed by a `## Table of Contents` listing each section with a 1-indexed line range and a short description (e.g. `- L48–65: Early history — Founded 1874 as the Nautical School...`), then a literal `=== DOCUMENT BODY ===` line, then the body with `## <heading>` markers at the listed line ranges.

How to read these docs:
(1) Start with `read(path, offset=0, limit=60)` to see the YAML + TOC + sentinel. If the result does NOT contain the `=== DOCUMENT BODY ===` line, the TOC is longer — re-read with a larger limit (e.g. `offset=0, limit=200`).
(2) The TOC is a navigation index, NOT an answer source. Each TOC entry is a short summary line, not the actual text — it can omit details, drop nuance, or be slightly off. Never extract a specific fact (number, name, date, quote) from a TOC summary line.
(3) Pick the most relevant section(s) from the TOC and read each one in full: `read(path, offset=<start-1>, limit=<end-start+1>)`. Confirm the answer in the actual section body before committing. If the body doesn't contain the expected fact, the TOC summary was misleading — read a wider chunk or another section.
(4) If multiple sections look plausible, read each. If no section looks right, read more of the doc (larger limit, or a different offset). Don't commit just because the TOC line sounds relevant.
(5) If the TOC says `(no sections — short or single-topic document; full text follows)`, read the file normally.

Use the tools iteratively: search to pull evidence into your working directory, bash/read to inspect it, and repeat until confident. When confident, output the final answer in this format and stop calling tools:
Explanation: {your reasoning}
Exact Answer: {your succinct final answer}
Confidence: {0-100%}
\end{lstlisting}

\subsection{Example Structured Document}
\label{app:structured_example}

The following abridged example is taken from the structured 100k corpus.
Its original corpus id is \texttt{2264}, and its corpus-relative path is
\texttt{www.bathspa.ac.uk/Our History.txt}. The excerpt shows the
inserted TOC, body sentinel, and several body spans; omitted body text is
marked explicitly.

\begin{lstlisting}[style=promptstyle]
---
title: Our History
date: 2025-01-01
---

## Table of Contents

- L18-23: Heritage and earliest roots — Introduces Bath Spa University's long heritage, then frames its origins in art education dating back over 170 years.
- L24-27: Bath School of Art era — Covers the Bath School of Art's 1852 establishment after the Great Exhibition and the school's reputation through named artists.
- L28-31: Postwar modernism and Corsham Court — Describes the Bath Academy of Art's postwar relocation to Corsham Court and its role in the English Modernist movement.
- L32-35: 1983 Bath Academy merger move — Explains the 1983 incorporation into Bath College of Higher Education and the move to Sion Hill alongside Newton Park's domestic science.
- L36-40: Newton Park and Sion Hill — Details the 1940s teacher-training origins, Joseph Langton's Grade 1 Newton Park Main House, and the Sion Hill opening.
- L41-44: 1975 merger and teacher education — Covers the 1975 merger into the Bath College of Higher Education and highlights the School of Education's South West teacher-training role.
- L45-47: Locksbrook Campus expansion 2019-2020 — Traces the 2016 purchase, Grimshaw-led refurbishment, the October 2019 Locksbrook Campus opening, and Jeremy Irons's 2020 formal opening.

=== DOCUMENT BODY ===

## Heritage and earliest roots

Our History

The University builds on a proud heritage, with our origins dating back over 170 years.

## Bath School of Art era

The story of Bath Spa University begins in 1852, when the Bath School of Art was established after the Great Exhibition. The School enjoyed a reputation as one of the leading art schools in the country, and some of Britain's best known artists, including Walter Sickert and Howard Hodgkin, studied and taught at what was to become the Bath Academy of Art.

[... body sections omitted ...]

## Locksbrook Campus expansion 2019-2020

In 2016 the University purchased the former Herman Miller factory, a listed building designed by Sir Nicholas Grimshaw in 1976, in Locksbrook Road on the River Avon. Remodelling and refurbishment was again designed by Grimshaw Architects, and the building was opened as the Locksbrook Campus in October 2019.
\end{lstlisting}

\section{Structured Corpus Generation Prompt}
\label{app:structured_generation_prompt}

The following prompt is used offline to propose section boundaries for
the structured corpus. The placeholder \texttt{<<<DOC>>>} is replaced by
the document text, possibly truncated as described in \Cref{sec:method}.

\begin{lstlisting}[style=promptstyle]
You restructure a plain-text document so a search agent (using bash tools `cat`, `sed`, `grep`) can navigate it.

You will NOT rewrite the document. You only propose:
  - section boundaries (where each section begins),
  - a short heading for each section, and
  - a one-sentence description of what each section covers.

A downstream script will use your output to insert `## <title>` headings into the ORIGINAL text and build a table-of-contents (with line numbers + your descriptions). The script locates each section by searching the document for the anchor string you provide. So the anchor is a SHORT LOCATOR — not the section's content.

OUTPUT FORMAT — return exactly this JSON, no commentary, no markdown fences:
{
  "sections": [
    {
      "title": "<short heading, 2-7 words, describing the section's content>",
      "anchor": "<the FIRST 6 to 12 WORDS of the section, copied EXACTLY from the document>",
      "description": "<one sentence, 10-25 words, summarizing what this section covers — designed for an agent skimming the TOC to decide if the section is worth reading>"
    },
    ...
  ]
}

ANCHOR RULES (CRITICAL — the script does exact substring matching):
- An anchor is ONLY a short locator string — typically 6 to 12 words. NEVER more than 12 words. Do not paste the whole paragraph.
- Copy those 6-12 words letter-for-letter from the document: same words, same punctuation, same capitalization, same internal whitespace.
- Each anchor must be a unique substring within the document.
- Anchors must appear in the same order as the sections appear in the document.
- The first section's anchor must be the first 6-12 words of the document BODY (skip the YAML `---` frontmatter at the top — do not anchor inside it).
- NEVER paraphrase, summarize, abbreviate, normalize, or "clean up" the anchor text. If the document says `Master of the SS United States`, the anchor must contain those exact words, not `Master of the` followed by a newline.
- If a section starts with a list, the anchor is the first 6-12 words of the FIRST list item (including any leading `*` or `-`), not multiple items.
- If you cannot locate a verbatim 6-12 word anchor in the document for a section you would like to propose, DO NOT propose that section. Under-sectioning is acceptable; inventing anchors is not.

SECTIONING GUIDELINES:
- Default behavior: ALMOST EVERY DOCUMENT HAS STRUCTURE WORTH MARKING. If the document has more than ~1500 characters of content, it almost certainly has 2+ sections you can identify (e.g., introduction + main content + conclusion, or topic A vs topic B).
- Only return {"sections": []} for genuinely short single-paragraph documents (under ~1000 characters of body text) OR documents that are pure metadata/boilerplate. When in doubt, propose sections — under-sectioning helps no one, but over-sectioning is mildly wasteful at worst.
- Split at natural topic shifts: a new subject, a clear temporal jump, a sub-topic the reader would expect a heading for.
- Default cadence for prose: one section per ~500-2000 words. A 3000-word narrative doc yields 2-6 sections, not 15.
- CATALOG / LIST documents are different and SHOULD be sectioned aggressively — one section per item is often correct:
  * A wildlife-identification guide listing 24 bird species → 24 sections (one per species).
  * A "list of films released in 2020" → either one big "Films" section, or one section per studio/country/letter grouping. Not 0 sections.
  * A Wikipedia franchise/series article with infobox + many distinct sub-topics → many sections.
  * A scientific paper with Abstract/Methods/Results/Discussion → at least those 4 sections.
- Procedural step lists (recipe instructions, how-to numbered steps, tutorials) belong to ONE umbrella section ("Instructions"), not one section per step. This is the OPPOSITE of catalog handling — steps are dependent and sequential, catalog items are independent.
- Section titles describe the content (e.g., "Early career at Bell Labs"), not generic ("Section 1", "Introduction" — those are too vague).
- Do NOT create a section for the YAML frontmatter itself.
- Do NOT create a section for trailing boilerplate (navigation, "Related links", "Follow us") unless it contains substantial content.

DESCRIPTION RULES:
- 10-25 words. Concrete, not generic. Mention key entities (names, dates, places, numbers) when those are the section's focus.
- Bad:  "Discusses the topic in more detail."
- Good: "Erling Persson's 1947 vision, opening of first womenswear store in Vasteras, family expansion through the 1950s."

DOCUMENT:
<<<DOC>>>
\end{lstlisting}

\end{document}